\documentclass[11pt,a4paper]{article}
\usepackage{float}
\usepackage{epsfig}
\usepackage{multirow}
\usepackage[cp1252,utf8]{inputenc}
\usepackage{hyperref}
\usepackage{graphicx} 
\usepackage{amsmath}
\usepackage{amsfonts}
\usepackage{amssymb}
\usepackage{bigints} 
\usepackage[top=2cm, bottom=2.5cm, left=1.5cm, right=1.5cm]{geometry}
\usepackage{authblk}
\usepackage{changepage} 
\usepackage{multicol}
\usepackage{afterpage}
\usepackage[normalem]{ulem}
\usepackage{placeins}

\usepackage{ragged2e}
\usepackage{graphicx,wrapfig}
\usepackage{bm}
\usepackage{hyperref}
\usepackage{mathrsfs}
\usepackage{subfig}
\usepackage{breqn}
\usepackage{caption}

\usepackage[english]{babel}

\title{\bf Shadow of Kottler black hole in the presence of plasma for a co-moving observer}
\author[a]{\bf Anish Das\thanks{anishdas1995@bose.res.in}}
\author[b]{\bf  Ashis Saha \thanks{ashisphys18@klyuniv.ac.in}}
\author[c]{\bf Sunandan Gangopadhyay \thanks{sunandan.gangopadhyay@bose.res.in}}

\affil[a,c]{\textit{Department of Astrophysics and High Energy physics,\linebreak
S.N.~Bose National Centre for Basic Sciences,}\linebreak
\textit{JD Block, Sector-III, Salt Lake, Kolkata 700106, India}}

\affil[b]{\textit{Department of Physics, University of Kalyani, \linebreak
Kalyani 741235, India}}

\date{}

\begin{document}
\maketitle

\begin{abstract}

\noindent In this paper, we discussed the shadow of a static, spherically symmetric black hole geometry in presence of a positive cosmological constant $\Lambda >0$. We analysed the black hole shadow both in absence and presence of plasma. Then we study the black hole shadow from the point of view of a comoving observer. We display the plots for the angular shadow size as measured by a comoving observer. The effects of the cosmological constant $\Lambda$ and plasma parameter $k$ on the angular size of the black hole shadow have been investigated in detail. Finally, by using the observed angular size of M87$^*$  and Sgr A$^*$ black hole, we constrain the value of the plasma parameter $k$ with a specific observational value of the cosmological constant $\Lambda$ (or the Hubble constant $H_0$).

\end{abstract}

\maketitle


\section{Introduction}\label{sec0}

\noindent Black holes are one of the most interesting predictions of general theory of relativity. Recent detection of black hole mergers by LIGO and Virgo in 2015 \cite{2} and the images of black hole shadow published by the EHT in 2019 \cite{3} and 2021 \cite{4} significantly triggered research in this direction. Here we take interest in studying the black hole shadow, which was initiated by the works of Synge \cite{5}. Synge calculated the \textit{escape cones of light} in case of Schwarzschild black hole. The escape cones basically reflect the angular portion where light can escape. In other words, they complement the black hole shadow, as we know it. The shadow for the spherically symmetric Schwarzschild black hole is circular. In general as far as we know most of the astrophysical objects rotate. So the general case is to consider a rotating black hole, the simplest of which is the Kerr black hole. The shadow in case of Kerr black hole was calculated by Bardeen in \cite{6}. In this case, the shadow is found to be deformed rather than circular due to presence of rotation. The reason being light gets dragged around the black hole due to rotation. Bardeen calculated the shadow from the point of view of an observer placed at infinity. In general the observer resides far away from the black hole, so the approximation of the observer being at infinity is quite logical. However while we need to compare our analytically computed results with the observations (like EHT data), we need to place the observer at a finite distance from the black hole. The computation of black hole shadow for a finite distant observer was first performed in \cite{7} and later in \cite{8} where they consider the presence of plasma. Further the black hole shadows were calculated by incorporating additional fields and hairs \cite{9}-\cite{6b} and with modified gravity theories in \cite{19}-\cite{36}. {Also, there are some recent interesting works in this direction which introduces new techniques to calculate photon spheres and black hole shadows \cite{n1}-\cite{n3}. Besides, there is a recent review work on analytical studies of black hole shadows \cite{n4}}.\\
It is a well-known fact that the universe is expanding and so any two objects in the universe are moving apart from one another. Although for nearby objects the effect is either less pronounced or superseded by gravitational attraction, still the effect persists. So from theoretical perspective inclusion of the cosmic expansion in the context of black hole shadow is quite important. The reason is that this would address the question of how does the black hole shadow depend on time. Furthermore, it can be readily mentioned that the size of any object increases due to cosmological expansion \cite{37}, \cite{38}. Inspired by this we consider a system which includes the cosmological expansion. The criteria that this system needs to satisfy is that it must be Schwarzschild like at nearby regions and for far away regions, it must reproduce the FLRW metric. The first of such model was considered by Einstein and Strauss \cite{39}. In this model, one calculates the geodesics in different regions and match them at a specific radius, namely the Sch\"ucking radius \cite{40}. On the other hand, the McVittie metric \cite{41} efficiently interpolates between the two mentioned regions. It satisfies the criteria of being Schwarzschild like in small distances and FLRW like at far distance. However the geodesics are not completely integrable analytically. Thus we must search for metrics where there exists sufficient constants of motion and the equations become completely integrable. A special case is that of Kottler or Schwarzschild de-Sitter metric as given in \cite{42}. The Kottler metric is a solution of Einstein equation with a positive cosmological constant $\Lambda > 0$. Here the system basically describes a Schwarzschild black hole being embedded in a de-Sitter universe. In this case, the geodesics are completely integrable and thus we can proceed to perform analytical computations. Besides a general transformation as proposed in \cite{1} can transform the Kottler metric to the FLRW metric where the expansion is driven by the cosmological constant $\Lambda$. Some recent works related to the case of expanding universe can be found in \cite{43}-\cite{k1}.\\
\noindent Another thing to be noted is that the black hole is in general surrounded by material media. The material around a black hole is at a very high temperature due to the immense gravitational field of the black hole. The nature of this material medium suggests that one can treat it like a highly dense plasma. Besides, the importance of considering a plasma medium lies in the fact that in a practical scenario, black holes are surrounded by a dispersive matter medium. Keeping this motivation in mind several studies have been done \cite{p1}-\cite{p4}. This plasma medium can either be magnetic or non-magnetic. The analysis of magnetic plasma is complicated analytically. So for the sake of simplicity, we restrict ourselves to non-magnetised plasma where the material is dust like with pressure $P=0$.  Numerous works on black hole shadow considering plasma can be found in literature. We list a few of them here \cite{50}-\cite{60d}. In our analysis, we aim to study black hole shadow in presence of plasma for a static and co-moving observer. We would like to study the impact of cosmological constant $\Lambda$ as well as the plasma parameter $k$ on the photon sphere as well as black hole shadow from the point of view of a co-moving observer moving away from the black hole due to cosmic expansion. We try to constraint the value of the plasma parameter on the basis of the observational results of M87$^*$ and Sgr A$^*$ black holes. Despite of the fact that the supermassive black holes are rotating still the comparison of the data with a non-rotating black hole can give us crude results which will help us get some insight into the plasma medium around the black hole.

\noindent The paper is organised as follows. In section \ref{sec0}, we gave an introduction and overview of the works on black hole shadow and reason for indulging in this work. In section \ref{sec1}, we discuss black hole shadow in presence of plasma. In section \ref{sec2}, we study the shadow of black hole with cosmological constant in presence of plasma. Then, in section \ref{sec3}, we discuss the black hole shadow from the point of view of a co-moving observer. Then, in section \ref{sec5}, we compare our results with observation in order to constrain the plasma parameter. Finally, we conclude in section \ref{sec6}. We also have an Appendix. We use geometric units for our calculation that is $c=G=\hbar=1$.

\section{Black hole shadow in presence of plasma}\label{sec1}

We start with a brief review of black hole shadow in the presence of plasma. The metric of a static, spherically symmetric black hole spacetime in $(3+1)$-dimensions is given by
\begin{equation}\label{m1}
ds^2 = -f(r)dt^2 + \frac{dr^2}{f(r)} + r^2 d \theta^2 + r^2 \sin ^2 \theta d \phi^2~.
\end{equation}  
To approach the problem, we first need to analyse the spacetime itself. The lapse function goes to zero at the event horizon $r= r_{h+}$, that is, $\Big(f(r)\Big|_{r=r_{h+}}=0\Big)$. This depends upon the spacetime parameters. The observer resides outside the event horizon. The black hole shadow boundary is formed by the photons that encircle the black hole. To be precise, the boundary is formed by light rays that are unstable which can either fall into the black hole or escape to infinity upon encountering any small radial perturbation. In order to determine the black hole shadow, we need to evaluate the geodesics of light rays by using a Lagrangian ($\mathcal{L}$) or Hamiltonian ($\mathcal{H}$). In general due to spherical symmetry, all planes are identical, so we wish to fix the plane of choice as $\theta=\frac{\pi}{2}$. Thus the geodesic for $\theta$ becomes $\theta=\frac{\pi}{2}=$constant or $\dot{\theta}=0$. Next we move on to calculate the other geodesics for $t$, $\phi$ and $r$. For this we use the Hamilton's equations of motion
\begin{equation}\label{23}
\dot{x}^{\mu}=\frac{\partial \mathcal{H}}{\partial p_{\mu}}~~;~~ \dot{p}_{\mu}=-\frac{\partial \mathcal{H}}{\partial x^{\mu}}~.
\end{equation}
\noindent We want to continue our discussion further by considering plasma. The reason being incorporation of plasma will make the case realistic. The plasma medium can be magnetised or non-magnetised. Consideration of magnetised plasma will resist us from performing an analytical study and we have to take help of numerical methods. So in order to study the system analytically we consider cold, dust-like ($P=0$) and non-magnetised plasma medium which corresponds to a Hamiltonian ($\mathcal{H}$) of the following form \cite{61}

\begin{equation}\label{35}
\mathcal{H}=\frac{1}{2}\Bigg[g^{\mu \nu}p_{\mu}p_{\nu} + \omega_p ^2\Bigg]~
\end{equation}
where $\omega_p$ represents the electron plasma frequency. The refractive index ($n$) of the plasma medium depends on the plasma frequency $\omega_p$ as well as the photon frequency ($\omega$), as measured by any arbitrary observer and the relation between them stands to be \cite{52}

\begin{equation}\label{36}
n^2= 1- \Bigg(\frac{\omega_p }{\omega}\Bigg)^2~~;~~\omega=\frac{\omega_0}{\sqrt{-g_{00}}}
\end{equation}
where the second equation gives the gravitational redshift relation with $\omega_0$ being the frequency of the photons as measured by a stationary observer at infinity. By setting $n = 1$, we get back the case of spherically symmetric spacetime devoid
	of plasma. Since the metric given in eq.\eqref{m1} is diagonal so its inverse can be computed easily. We can write
\begin{equation}
\omega=\frac{\omega_0}{\sqrt{-g_{00}}}~~ \Rightarrow~~ \hbar\omega=\frac{\hbar\omega_0}{\sqrt{-g_{00}}}~~ \Rightarrow~~ \hbar\omega=\frac{E_0}{\sqrt{-g_{00}}}~~ \Rightarrow~~ \hbar\omega=\frac{p_0}{\sqrt{-g_{00}}}~~\Rightarrow~~ \hbar\omega=p_0\sqrt{-g^{00}}~~.
\end{equation}
For $\hbar =1 $, we have $\omega=p_0\sqrt{-g^{00}}$. Using this relation and eq.\eqref{36} in eq.\eqref{35}, we have

\begin{equation}
\mathcal{H}=\frac{1}{2}\Bigg[g^{\mu \nu}p_{\mu}p_{\nu} +(n^2 -1) g^{00}p_0^2 \Bigg]~.
\end{equation}
 
\noindent The Hamiltonian can be recast as
\begin{equation}
\mathcal{H}=\frac{1}{2}\Bigg[n^2 g^{00}p_{0}^2 + g^{ij}p_i p_j \Bigg]
\end{equation}
where $i, j$ runs from 1 to 3. The metric and hence the Hamiltonian $\mathcal{H}$ is independent of $t$ and $\phi$. Also by using the Hamilton's equation of motion $\dot{p}_{\mu}=-\frac{\partial \mathcal{H}}{\partial x^{\mu}}$ and using the fact that $\omega_p =\omega_p (r)$, we get the constants of motion as $p_0=$constant$=-E$ and $p_3 =$constant$=L$. Here $E$ and $L$ correspond to the energy and angular momentum of photons as measured by an observer stationed at infinity.  Using the Hamilton's equations of motion given in eq.\eqref{23} and by restricting ourselves to the equatorial plane, we get the geodesics for $t$, $\phi$ and $r$
as
\begin{equation}
\frac{d t}{d \lambda}=\frac{n^2 E}{f(r)}
\end{equation}
\begin{equation}\label{38}
\frac{d \phi}{d \lambda}=\frac{L}{r^2}
\end{equation}
\begin{equation}\label{39}
\Big(\frac{d r}{d \lambda}\Big)^2=n^2 E^2 -\frac{L^2}{r^2}f(r)~.
\end{equation}

Here $\lambda$ corresponds to the affine parameter. Taking the ratio of eq.(s)\eqref{38} and \eqref{39}, we have
\begin{equation}\label{40}
\Big(\frac{d r}{d \phi}\Big)^2 = r^4 \Bigg[\frac{n^2 E^2}{L^2}- \frac{f(r)}{r^2}\Bigg]~.
\end{equation}

\noindent The solution of the above equation gives the trajectory of photons as $r=r(\phi)$ in presence of plasma. Eq.\eqref{40} can be used to evaluate the unstable photon orbits by imposing the conditions $\frac{d r}{d \phi}=\frac{d^2 r}{d \phi^2}=0$.

\noindent The first condition constraints the constants of motion ($E$, $L$) in terms of the photon sphere radius $r_p$ and the second condition helps to determine the photon sphere radius $r_p$ in terms of spacetime parameters. The conditions take the following form
\begin{eqnarray}
\frac{n^2 r ^2}{f(r)}\Bigg|_{r=r_p}&=&\frac{L^2}{E^2} \\
\Big(2 n' r f(r) + 2nf(r)-n r f'(r)\Big)\Bigg|_{r=r_p}&=&0~.
\end{eqnarray}

\begin{wrapfigure}{l}{0.5\textwidth}
\begin{center}
\includegraphics[width=8.5cm]{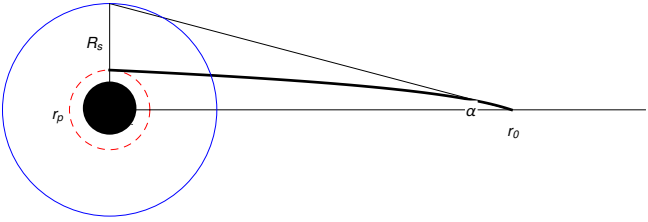}
\end{center}
\caption{ \footnotesize Pictorial representation of the event horizon, photon sphere and the black hole shadow along with the trajectory of light ray.}
\label{x1}
\end{wrapfigure}

\noindent The second equation gives a fixed value of $r_p$ for fixed values of spacetime parameters and the refractive index $n$ thereby giving a constant value for $\frac{L^2}{E^2}$. So we see that along a photon trajectory $\frac{L^2}{E^2}$ is constant.\\
Next we wish to calculate the angular size of the black hole shadow. A ray of light from the background source upon getting deflected due to the immense gravitational pull of the black hole, makes an angle $\alpha$ with respect to the observer's position $r_0$ as given in Fig.\ref{x1}. The Figure shows a dark disk which represents the black hole being surrounded by photon sphere (shown in dotted red colour) situated at the distance $r=r_p$. Light from photon sphere travels in curved trajectory and reaches the observer at an angle $\alpha$ where the tangent gives the measured position of the light ray. If we consider all rays at the boundary of the cone of angle $\alpha$, it forms the shadow boundary with radius $R_s$ greater than $r_p$, that is $R_s > r_p > r_{h+}$.

\noindent In the above, we have assumed the plasma distribution to have only radial dependence ($n=n(r)$) and $n'=\frac{dn}{dr}$. This is due to the presence of spherical symmetry.
In order to carry out calculations explicitly, we need to assume a certain form of the plasma frequency ($\omega_p$) and thereby refractive index ($n$). Keeping this in mind, we consider the following form for the plasma frequency\footnote{It is to be noted that as we are working with an axially symmetric black hole it is justified to assume the form of $\omega_p$ as $\omega_p ^2=\frac{f(r)}{r^h}$. However, for a rotating black hole one needs to consider the form of $\omega_p$ as $\omega_p ^2=\frac{f(r)+f(\theta)}{r^2+a^2\cos\theta}$ as shown in \cite{60a}.} \cite{52}
\begin{equation}
\omega_p(r)^2 = \frac{4\pi e^2}{m_e}N(r)~~;~~ N(r)=\frac{N_0}{r^h}
\end{equation}
where $e$ and $m_e$ represents the electronic charge and mass, $N(r)$ gives the number density of electrons in the plasma medium and $N_0$ is a constant. Substitution of the above relation in the expression of refractive index $n$  in eq.\eqref{36} gives
\begin{equation}
n(r)^2= 1-f(r)\frac{k}{r^h}~~;~~k=\frac{4\pi e^2}{m_e \omega_0 ^2}N_0~.
\end{equation}
\begin{wrapfigure}{l}{0.35\textwidth}
\begin{center}
\includegraphics[width=6.5cm]{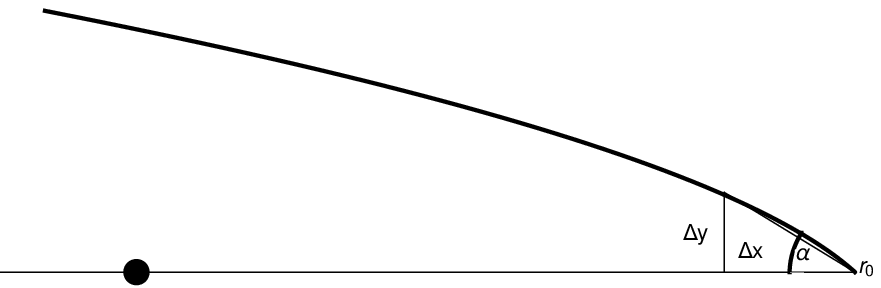}
\end{center}
\caption{ \footnotesize Light travelling into past from observer ($r_0$) making angle $\alpha$.}
\label{2}
\end{wrapfigure}
Next we try to evaluate the angular size of black hole shadow in presence of plasma. In Fig.\ref{2} we find that the ray of light makes an angle $\alpha$ with the radial line that satisfies
\begin{equation}
\tan \alpha = \lim_{\Delta x \to 0} \frac{\Delta y}{\Delta x}~.
\end{equation}
From our metric we find that in the desired limit, the angular size becomes
\begin{equation}\label{tan1}
\tan \alpha=\sqrt{\frac{f(r)}{r^2 \frac{n(r)^2 E^2}{L^2}-f(r)}}\Bigg|_{r=r_o}
\end{equation}
where $r_o$ is the position of the observer. The angular size of black hole shadow in terms of $\sin \alpha$ takes the form
\begin{equation}\label{14}
\sin^2 \alpha_{sp} = \frac{f(r)}{r^2 \frac{n(r)^2 E^2}{L^2}}\Bigg|_{r=r_o} =\frac{z(r_p)^2}{z(r_0)^2}~~;~~z(r)^2=\frac{n(r)^2 r^2}{f(r)}~.
\end{equation}

\section{Shadow of black hole with a positive cosmological constant in plasma}\label{sec2}
We now consider the presence of a positive cosmological constant $\Lambda$.  The positive cosmological constant ($\Lambda > 0$) introduces an additional horizon, namely, the cosmological horizon ($r_C$). The cosmological horizon constraints the observers existence within a certain region from the event horizon $r_{h+}$ to cosmological horizon $r_C$. Also, analysing the lapse function $f(r)$ we can see that the position of the event horizon is shifted outward. The interesting thing we can observe is that the photon sphere radius $r_p$ remains unchanged implying no effect of cosmological constant $\Lambda$ \cite{43}. However the angular shadow size not only depends on the photon sphere radius $r_p$ but also explicitly depends on the specification of the geometry (the lapse function) of the black hole which in turn depends on the cosmological constant $\Lambda$. This observation can be verified from eq.\eqref{14}.

\noindent Now we wish to study the Schwarzschild de-Sitter metric or the Kottler metric. The metric is given as
\begin{equation}\label{m2}
ds^2 = -\Big(1-\frac{2M}{r} - \frac{\Lambda}{3} r^2\Big)dt^2 + \frac{dr^2}{\Big(1-\frac{2M}{r} - \frac{\Lambda}{3}r^2\Big)} + r^2 d \theta^2 + r^2 \sin ^2 \theta d \phi^2
\end{equation}
with the cosmological constant $\Lambda$ related to the Hubble's constant $H_0$ as $\frac{\Lambda}{3}=\frac{H_0 ^2}{c^2}$. From the lapse function $f(r)$ we observe that the horizons lie in the ranges $2M<r_{h+}<3M$, $3M<r_C<\infty$ and $\bar{r}<0$, where $r_{h+}$, $r_C$ and $\bar{r}$ respectively give the event horizon, cosmological horizon and the additional horizon which is unphysical. Also setting $f(r)=0$ provides us a cubic equation of the form
\begin{equation}
r^3 - \frac{3}{\Lambda}r +\frac{6M}{\Lambda}=0
\end{equation}
which constraints $\Lambda$ as $0< \Lambda < \frac{1}{9M^2}$. The geodesics are evaluated at the equatorial plane ($\theta=\frac{\pi}{2}$) by incorporating plasma. The geodesics for $t$, $\phi$ and $r$ take the following forms

\begin{equation}
\frac{d t}{d \lambda}=\frac{n^2 E}{f(r)}=\frac{n^2 E}{\Big(1-\frac{2M}{r} - \frac{\Lambda}{3} r^2\Big)}
\end{equation}
\begin{equation}
\frac{d \phi}{d \lambda}=\frac{L}{r^2}
\end{equation}
\begin{equation}
\Big(\frac{d r}{d \lambda}\Big)^2=n^2 E^2 -\frac{L^2}{r^2}f(r)=n^2 E^2 -\frac{L^2}{r^2}\Big(1-\frac{2M}{r} - \frac{\Lambda}{3} r^2\Big)~.
\end{equation}
The condition for unstable circular null geodesics gives the photon sphere radius $r_p$ and also it gives a condition on the constants $E, L$ and $\Lambda$ as
\begin{equation}
\frac{L^2}{E^2}=\frac{n^2 r ^2}{f(r)}\Bigg|_{r=r_p}
\end{equation}
which simplifies to
\begin{equation}
\frac{E^2}{L^2}=\frac{1}{27M^2}-\frac{\Lambda}{3}
\end{equation}
for $n=1$ as was found in \cite{43}. The angular size of the black hole shadow in presence of plasma takes the form
\begin{equation}
\sin^2 \tilde{\alpha}_{sp} =\frac{z(r_p)^2}{z(r_0)^2}=\frac{n(r_p)^2}{n(r_0)^2}\frac{r_p ^2}{r_0 ^2}\frac{\Big(1-\frac{2M}{r_0} - \frac{\Lambda}{3} r_0 ^2\Big)}{\Big(1-\frac{2M}{r_p} - \frac{\Lambda}{3} r_p ^2\Big)}.
\end{equation}
The general equation for deriving the photon sphere radius $r_p$ for static spherically symmetric metric with cosmological constant $\Lambda$ and plasma medium with refractive index $n$ is given as
\begin{equation}\label{57}
\Big(2 n'(r) r f(r) + 2n(r)f(r)-n(r) r f'(r)\Big)\Bigg|_{r=r_p}=0~.
\end{equation}
Let us solve try to solve the equation for different cases. \\

\textbf{Case I}\\
Here we consider \textit{Schwarzschild de Sitter metric or Kottler metric without plasma}. The lapse function is of the form 
\begin{equation}
 f(r)=1-\frac{2M}{r}-\frac{\Lambda}{3}r^2   
\end{equation}
and the corresponding derivative is
\begin{equation}
  f'(r)=\frac{df}{dr}=\frac{2M}{r^2}-\frac{2\Lambda}{3}r~.  
\end{equation}
 Since there is no plasma, hence $n(r)=1$ and thereby $n'(r)=0$. Replacing in eq.\eqref{57}, we get
\begin{equation}
2\Big(1-\frac{2M}{r_p}-\frac{\Lambda}{3}r_p ^2\Big)-r_p\Big(\frac{2M}{r_p ^2}-\frac{2\Lambda}{3}r_p\Big)=0 \Rightarrow r_p=3M.
\end{equation}
So we find that the presence of cosmological constant $\Lambda$ in the metric does not effect the photon sphere radius $r_p$ in absence of plasma. The corresponding angular shadow size takes the form \\

\begin{equation}\label{56}
\sin^2 \tilde{\alpha}_s =\frac{r_p ^2}{r_0 ^2}\frac{\Big(1-\frac{2M}{r_0} - \frac{\Lambda}{3} r_0 ^2\Big)}{\Big(1-\frac{2M}{r_p} - \frac{\Lambda}{3} r_p ^2\Big)} =\frac{\Big(1-\frac{2M}{r_0} - \frac{\Lambda}{3} r_0 ^2\Big)}{r_0^2 \Big(\frac{1}{27M^2}-\frac{\Lambda}{3}\Big)}~.
\end{equation}

\textbf{Case II}

\noindent Here we consider \textit{Schwarzschild de Sitter or Kottler metric with homogeneous plasma}. The refractive index of homogeneous plasma is given as \cite{p2}, \cite{55}
\begin{equation}
    n(r)=\sqrt{1-kf(r)}
\end{equation}
and the corresponding derivative 
\begin{equation}
    n'(r)= -\frac{1}{\sqrt{1-kf(r)}}k\Bigg(\frac{M}{r^2}-\frac{\Lambda}{3}r^2\Bigg)~.
\end{equation}
Replacing in eq.\eqref{57}, we get after simplifying
\begin{equation}\label{p1}
\frac{k \Lambda^2}{9}r_p ^6 - \frac{2}{3}\Lambda k r_p ^4 + \frac{4k \Lambda M}{3}r_p ^3 -(1-k)r_p ^2 - (4k-3)Mr_p + 4kM^2=0.  
\end{equation}
The above equation is of sixth order. The solution gives the photon sphere radius $r_p$ which is a function of $k$ and $\Lambda$. In the limit of $k \to 0$, we get back the photon sphere radius $r_p$ as 3M. The angular shadow size takes the form\\

\begin{equation}\label{62}
\sin^2 \tilde{\alpha}_{sp} =\frac{n(r_p)^2}{n(r_0)^2}\frac{r_p ^2}{r_0 ^2}\frac{\Big(1-\frac{2M}{r_0} - \frac{\Lambda}{3} r_0 ^2\Big)}{\Big(1-\frac{2M}{r_p} - \frac{\Lambda}{3} r_p ^2\Big)}=\frac{\Bigg(1-k\Bigg(1-\frac{2M}{r_p}-\frac{\Lambda}{3}r_p ^2\Bigg)\Bigg)}{\Bigg(1-k\Bigg(1-\frac{2M}{r_0}-\frac{\Lambda}{3}r_0 ^2\Bigg)\Bigg)}\frac{r_p ^2}{r_0 ^2}\frac{\Big(1-\frac{2M}{r_0} - \frac{\Lambda}{3} r_0 ^2\Big)}{\Big(1-\frac{2M}{r_p} - \frac{\Lambda}{3} r_p ^2\Big)}.
\end{equation}

\textbf{Case III}

\noindent Here we consider \textit{Schwarzschild de Sitter or Kottler metric with inhomogeneous plasma}. The refractive index of inhomogeneous plasma is given as \cite{p2}, \cite{55}
\begin{equation}
     n(r)=\sqrt{1-\frac{k}{r}f(r)}
\end{equation}
and the corresponding derivative 
\begin{equation}
    n ' (r)=\frac{1}{2 \sqrt{1-\frac{k}{r}f(r)}}\Bigg(\frac{k}{r^2}-\frac{2kM}{r^3}-\frac{\Lambda k}{3}-\frac{2kM}{r^3}
+\frac{2}{3}k\Lambda\Bigg)~.
\end{equation}
Replacing in eq.\eqref{57}, we get after simplifying
\begin{equation}\label{p2}
\frac{k \Lambda^2}{9}r_p ^6 - \frac{2}{3}\Lambda k r_p ^4 + \Bigg(\frac{4k \Lambda M}{3}-2\Bigg)r_p ^3 + (6M + k)r_p ^2 - 4kr_p M + 4kM^2=0.    
\end{equation}
The above equation is of sixth order. The solution gives the photon sphere radius $r_p$ which is a function of $k$ and $\Lambda$. In the limit of $k \to 0$, we get back the photon sphere radius $r_p$ as 3M. The angular shadow size takes the form\\

\begin{equation}\label{64}
\sin^2 \tilde{\alpha}_{sp} =\frac{n(r_p)^2}{n(r_0)^2}\frac{r_p ^2}{r_0 ^2}\frac{\Big(1-\frac{2M}{r_0} - \frac{\Lambda}{3} r_0 ^2\Big)}{\Big(1-\frac{2M}{r_p} - \frac{\Lambda}{3} r_p ^2\Big)}=\frac{\Bigg(1-\frac{k}{r_p}\Bigg(1-\frac{2M}{r_p}-\frac{\Lambda}{3}r_p ^2\Bigg)\Bigg)}{\Bigg(1-\frac{k}{r_0}\Bigg(1-\frac{2M}{r_0}-\frac{\Lambda}{3}r_0 ^2\Bigg)\Bigg)}\frac{r_p ^2}{r_0 ^2}\frac{\Big(1-\frac{2M}{r_0} - \frac{\Lambda}{3} r_0 ^2\Big)}{\Big(1-\frac{2M}{r_p} - \frac{\Lambda}{3} r_p ^2\Big)}.
\end{equation}

\begin{figure}[H]
	\centering
	\begin{minipage}[b]{0.4\textwidth}
		\subfloat[\footnotesize Variation of photon sphere radius $\frac{r_p}{M}$ with restpect to the plasma parameter $k$ (homogeneous plasma). The plots are shown for different values of cosmological constant $\Lambda M^2$ as 0.0075 (black), 0.0300 (red) and 0.0675 (blue). ]{\includegraphics[width=\textwidth]{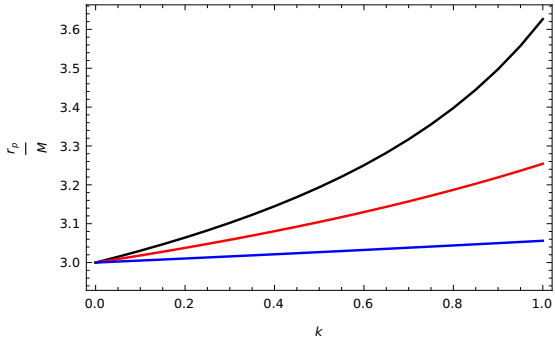}}
	\end{minipage}
	\hspace{1.0cm}
	\begin{minipage}[b]{0.4\textwidth}
		\subfloat[\footnotesize  Variation of photon sphere radius $\frac{r_p}{M}$ with respect to the plasma parameter $\frac{k}{M}$ (inhomogeneous plasma). The plots are shown for different values of cosmological constant $\Lambda M^2$ as 0.0075 (black), 0.0300 (red) and 0.0675 (blue). ]{\includegraphics[width=\textwidth]{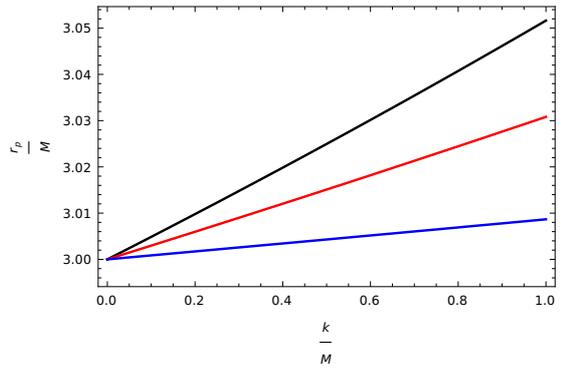}}
	\end{minipage}
	\caption{Variation of photon sphere radius $\frac{r_p}{M}$ with plasma parameter.}
	\label{89}
\end{figure}

\noindent In Fig.\ref{89}, we show the variation of the photon sphere radius $r_p$ with respect to the plasma parameter $k$. The left plot is for homogeneous plasma and the right one is for inhomogeneous plasma. In both the cases, we find that the photon sphere radius $r_p$ increases with increase in plasma parameter though the effect is more pronounced in case of homogeneous plasma. Further, we have chosen three different values of the cosmological constant $\Lambda$ for both of the plots. We find that with increase in the value of the cosmological constant, the value of the photon sphere radius decreases for a fixed value of $k$. Also, we find that all the plots start from the same common point at $k=0$ implying the fact that in absence of plasma, the cosmological constant has no effect on the photon sphere radius $r_p$. This observation can also be verified from eq.(s) \eqref{p1} and \eqref{p2}.\\

\begin{figure}[H]
\centering
\begin{minipage}[b]{0.4\textwidth}
\subfloat[\footnotesize Angular size of the shadow ($\alpha_{stat}$) of the Schwarzschild black hole, with respect to the position of a static observer ($r_0$). ]{\includegraphics[width=\textwidth]{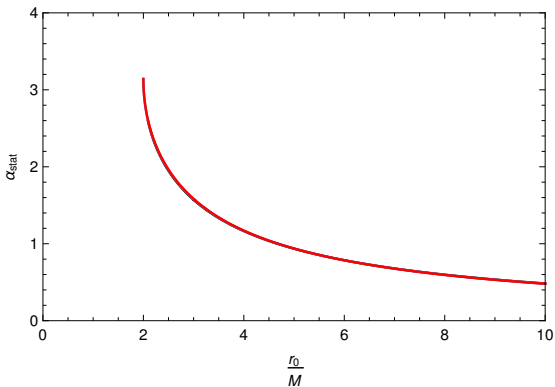}}\label{3a1}
\end{minipage}
\hspace{1.0cm}
\begin{minipage}[b]{0.4\textwidth}
\subfloat[\footnotesize  Angular size of the shadow ($\alpha_{stat}$) of the Schwarzschild de-Sitter black hole, with respect to the position of a static observer ($r_0$). We have set the value of positive cosmological constant $\Lambda M^2 =0.03$. ]{\includegraphics[width=\textwidth]{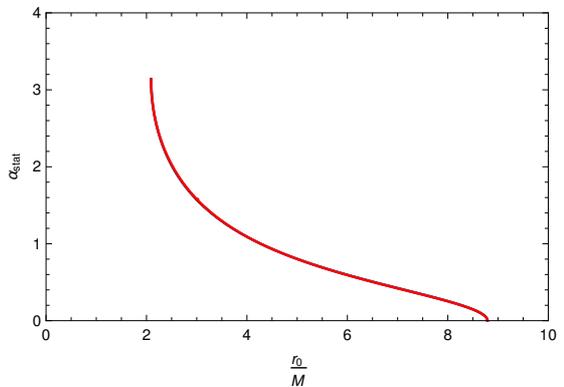}}\label{4a1}
\end{minipage}
\caption{Variation in angular size of the shadow with respect to the position of the observer $r_0$. The plots are shown for different values of the plasma parameter $\frac{k}{M}$ as 0.0 (black), 0.2 (violet) and 0.4 (red).}
\label{81}
\end{figure}
\begin{figure}[H]
\centering
\begin{minipage}[b]{0.4\textwidth}
\subfloat[\footnotesize Angular size of the shadow ($\alpha_{stat}$) with respect to the position of a static observer ($r_0$) for Schwarzschild black hole (zoomed).  ]{\includegraphics[width=\textwidth]{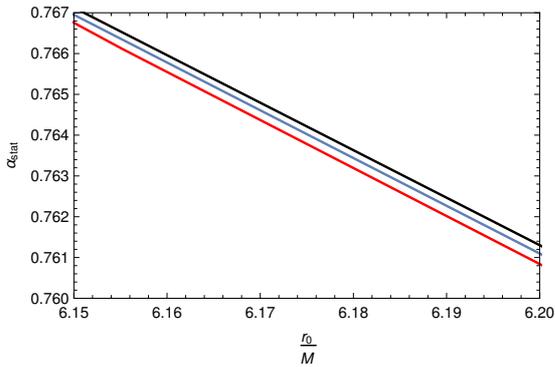}}\label{3a21}
\end{minipage}
\hspace{1.0cm}
\begin{minipage}[b]{0.4\textwidth}
\subfloat[\footnotesize  Angular size of the shadow ($\alpha_{stat}$) with respect to the position of a static observer ($r_0$) for Schwarzschild de-Sitter black hole with $\Lambda M^2 =0.03$ (zoomed). ]{\includegraphics[width=\textwidth]{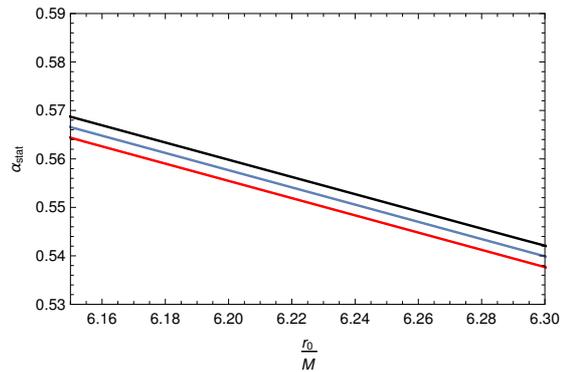}}\label{4a21}
\end{minipage}
\caption{Variation in angular size of the shadow with respect to the position of the observer $r_0$. The plots are shown for different values of plasma parameter $\frac{k}{M}$ as 0.0 (black), 0.2 (violet) and 0.4 (red).}
\label{821}
\end{figure}
\noindent In Fig.\ref{81}, we show the variation of the angular size ($\alpha$) of the black hole shadow. The plots are shown for different values of inhomogeneous plasma parameter $\frac{k}{M}$. We calculated the angular shadow size in terms of $\sin \alpha$. 

 \noindent As the observer moves from infinity (in case of an asymptotically flat black hole) or from the cosmological horizon $r_C$ (in case of an asymptotically de-Sitter black hole) towards the black hole, the angular shadow size starts to increase from 0 to $\frac{\pi}{2}$ and then falls back to zero. However, to plot the angular shadow in the range $\alpha \in [0,\pi]$, we shall take $\alpha = \sin^{-1} Q$ in the range $r_p < r_0 < r_C$ and $\alpha = \pi - \sin^{-1} Q $ in the range $r_h < r_0 < r_p$ \cite{43}. Here $Q$ corresponds to the value of $\sin \alpha$ as a function of radial distance $r_0$. \\  
  The left plot in Fig.\ref{81} is for Schwarzschild black hole where we found that the position of the observer can extend upto infinity. On the other hand, the right plot in Fig.\ref{81} shows the variation of the angular shadow size with observer distance $r_0$ in case of a Schwarzschild black hole embedded in a de-Sitter universe. Due to this embedding, the observer can reach upto a certain limiting radius, that is the cosmological horizon $r_0 = r_C$. In this case, the shadow size is zero at the cosmological horizon $r_C$ and increases to $\frac{\pi}{2}$ at the photon sphere $r_p$. After that as the observer comes closer to the black hole, the shadow size grows and becomes $\pi$ at the event horizon.

\noindent  The plots in Fig.\ref{821} are the zoomed plots of the plots in Fig.\ref{81}. They display the variation of the angular shadow size with variation in inhomogeneous  plasma parameter $\frac{k}{M}$ as observed by a distant observer. We find that the shadow size decreases with increase in plasma parameter $\frac{k}{M}$. Also we mention that the plots are shown for plasma frequency $\omega_p ^2$ having only radial dependence as $\omega_p ^2 \sim \frac{1}{r}$.  

\begin{figure}[H]
\centering
\begin{minipage}[b]{0.4\textwidth}
\subfloat[\footnotesize Angular size of the shadow ($\alpha_{stat}$) with respect to the position of a static observer ($\frac{r_0}{M}$) for Schwarzschild de-Sitter black hole with plasma parameter $\frac{k}{M}=0$.  ]{\includegraphics[width=\textwidth]{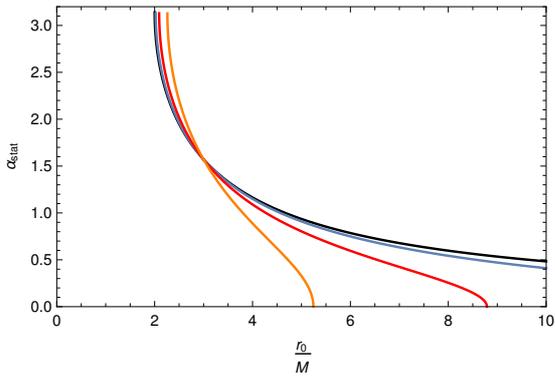}}\label{3a3}
\end{minipage}
\hspace{1.0cm}
\begin{minipage}[b]{0.4\textwidth}
\subfloat[\footnotesize  Angular size of the shadow ($\alpha_{stat}$) with respect to the position of a static observer ($\frac{r_0}{M}$) for Schwarzschild de-Sitter black hole with inhomogeneous plasma parameter $\frac{k}{M}=0.2$. ]{\includegraphics[width=\textwidth]{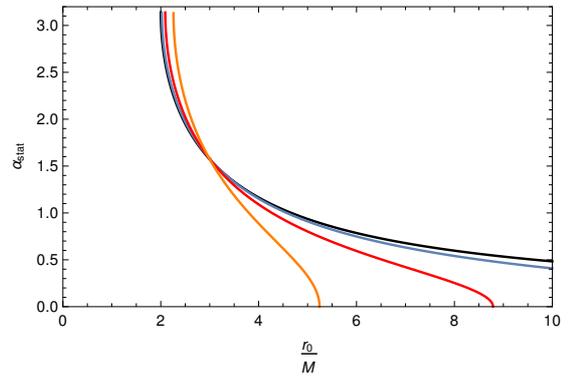}}\label{4a3}
\end{minipage}
\caption{Variation in the angular size of the shadow with respect to the position of the observer $\frac{r_0}{M}$. The plots are shown for different values of cosmological constant $\Lambda M^2$ as 0.0000 (black), 0.0075 (violet), 0.0300 (red) and 0.0675 (orange).}
\label{83}
\end{figure}
 \noindent The plots given in Fig.\ref{83} shows the variation of the black hole shadow for different values of the cosmological constant $\Lambda$ or in other words the Hubble's constant $H_0$. The plots are shown for Schwarzschild de-Sitter black hole with the values of the cosmological constant ($\Lambda M^2$) as 0.0000 (black), 0.0075 (violet), 0.0300 (red) and 0.0675 (orange). With increase in the value of $\Lambda$, we find that the shadow size decreases more quickly to zero as the cosmological horizon ($r_C$) comes closer to the black hole event horizon ($r_{h+}$). Besides, we also find that the black hole event horizon shifts outward with increase in the value of the cosmological constant $\Lambda$. Simply, we can say that $r_{h+}$ increases with increase in $\Lambda$ whereas $r_C$ decreases for the same and thereby they merge ($r_{h+}=r_C$) for $\Lambda M^2=0.1111$. The left plot is for inhomgeneous plasma parameter $\frac{k}{M}=0$ and the right plot is for $\frac{k}{M}=0.2$. The two plots are almost identical since the variation due to plasma gets reflected in the photon sphere radius $r_p$ but no effect of plasma is seen in case of $r_{h+}$ and $r_C$.

\section{Shadow from the point of view of co-moving observer}\label{sec3}
Next we try to study the effect of the cosmic expansion on the black hole shadow. The shadow is viewed by an observer stationed at a certain position in the domain of outer communication. The cosmological expansion drives the observer away from the black hole with time and is dictated by the cosmological constant $\Lambda$. The angular size of the shadow measured by this co-moving observer is related to that for a static observer by the aberration relation \cite{62}, \cite{63}
\begin{equation}
\cos \alpha_c = \frac{\cos \alpha_s - v}{1-v \cos \alpha_s}
\end{equation}
which in terms of sine function takes the form
\begin{equation}\label{66}
\sin ^2 \alpha_{c}=\Big(1-v^2\Big)\frac{\sin^2 \alpha_s}{\Big(1-v\cos \alpha_s\Big)^2}
\end{equation}
where $v$ is the relative velocity between the two observers. In general, we consider the co-moving observer is moving away from the static observer with velocity $v$. This velocity depends on the cosmological constant $\Lambda$ and thereby the Hubble constant $H_0$. Also it must be pointed out that as the static observer goes from $r_{h+}$ to $r_p$, the angular size of the black hole shadow changes from $\pi$ to $\frac{\pi}{2}$ and then goes from $\frac{\pi}{2}$ to 0 as the observer goes to $r_C$. This observation is clear from the right panel plot of Fig.\ref{81}. We have shown numerical values for the angles. We also point that in our case sine and cosine functions are allowed to take only positive values in the calculation, hence we put negative values by hand for a continuation of our results. Using this and by expressing $\cos \alpha$ in terms of $\sin \alpha$ eq.\eqref{66} becomes
\begin{equation}\label{60}
\sin \alpha_{c}=\sqrt{\Big(1-v^2\Big)}\frac{\sin \alpha_s}{\Big(1\pm v\sqrt{1- \sin^2 \alpha_s}\Big)}
\end{equation}
which can be rewritten as
\begin{equation}\label{v}
\sin \alpha_{c}=\sqrt{\Big(1-v^2\Big)}\sin \alpha_s \frac{\Big(1\mp v\sqrt{1- \sin^2 \alpha_s}\Big)}{\Big(1 - v^2 (1- \sin^2 \alpha_s)\Big)}~.
\end{equation}

\noindent The associated graphical representations show very interesting results. If we consider only the effect of the motion of the observer, we find that the shadow has certain features quite different from that of the static observer. In case of a static observer, as the observer moves away from the event horizon $r_{h+}$ the complete dark sky as seen by him starts to reduce in size and just at the photon sphere $r_p$, he sees half dark and half bright sky. Further, as he moves outward he eventually sees a completely bright sky. Also at $r_p$ light rays are perpendicular to the observer. But for the co-moving observer the situation is not the same.  For the comoving observer the effective photon sphere $\tilde{r}_p $ is somewhere outward, that is $\tilde{r}_p > r_p$. The rest portion is same as that of static observer. Also the position of the comoving observer is not bound by the cosmological horizon $r_C$ and the observer can readily reach upto infinity.

\noindent If we incorporate plasma now, the aberration relation gets modified to a great extent as given in \cite{64} as
\begin{equation}
\cos \alpha_{cp} = \frac{\cos \alpha_{sp} + nv}{\sqrt{\Big(n + v \cos \alpha_{sp}\Big)^2 - \Big(n^2 -1\Big)\Big(1-v^2\Big)}}
\end{equation}
which can be rewritten in terms of sine function as
\begin{equation}\label{20a}
\sin \alpha_{cp} = \sqrt{\Big(1-v^2\Big)}\frac{\sin \alpha_{sp}}{\sqrt{\Big(n \mp v \sqrt{1- \sin^2 \alpha_{sp}}\Big)^2 - \Big(n^2 -1\Big)\Big(1-v^2\Big)}}~.
\end{equation}
For $n=1$, we get back eq.\eqref{60}. The simplified form of the above eq.\eqref{20a} is given as
\begin{equation}\label{72}
\sin \alpha_{cp} =\sqrt{\Big(1-v^2\Big)}\sin \alpha_{sp}\frac{\sqrt{\Big(1+ n^2 v^2 -v^2 \sin ^2 \alpha\Big) \mp 2 n v \sqrt{\Big(1- \sin^2 \alpha\Big)}}}{\sqrt{\Big(1+ n^2 v^2 -v^2 \sin ^2 \alpha\Big)^2 - 4 n^2 v^2 \Big(1- \sin^2 \alpha\Big)}}~.
\end{equation}
Using this expression we can find the angular size of the black hole shadow surrounded by plasma as seen by a comoving observer moving away from the black hole.

\noindent Now, in order to plot the angular size of the black hole shadow, we need to determine the velocity of the comoving observer with respect to a static observer. The derivation has been performed by Perlick et al in \cite{43}. Also it has been done for the general case of any static and spherically symmetric spacetime by Roy et al in \cite{1}. The velocity of the comoving observer with respect to the static observer takes the form    

\begin{equation}
|\Vec{v}|=\sqrt{1-\frac{f(r)}{f_0(r)}}.
\end{equation}
with $f(r)$ being the lapse function and $f_0 (r) = f(r)\Big|_{\Lambda=0}$. In case of Kottler metric, the lapse function takes the form $f(r)=1-\frac{2M}{r}-\frac{\Lambda}{3}r^2$, which results into the velocity of comoving observer as
\begin{equation}\label{H}
|\Vec{v}|=\sqrt{\frac{\Lambda/3}{1-\frac{2M}{r}-\frac{\Lambda}{3}r^2}}r=\frac{H_0 r}{\sqrt{1-\frac{2M}{r}-H_0 ^2r^2}}.
\end{equation}
We have briefly discussed the formula for the velocity of the comoving observer with respect to the static observer in the Appendix.

\begin{figure}[H]
\centering
\begin{minipage}[b]{0.4\textwidth}
\subfloat[\footnotesize Angular size ($\alpha_{comov}$) of the Schwarzschild de-Sitter black hole shadow with respect to the position of a co-moving observer ($\frac{r_0}{M}$). We set the plasma parameter $\frac{k}{M}=0$.  ]{\includegraphics[width=\textwidth]{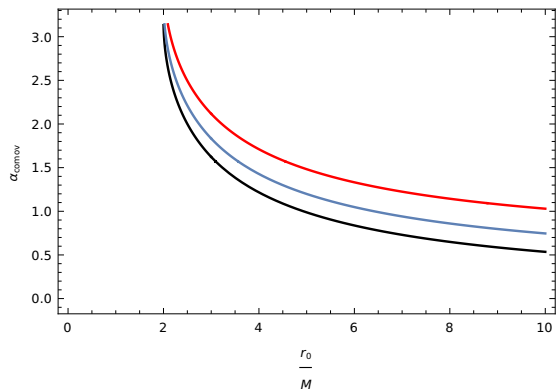}}\label{xx1}
\end{minipage}
\hspace{1.0cm}
\begin{minipage}[b]{0.4\textwidth}
\subfloat[\footnotesize  Angular size ($\alpha_{comov}$) of the Schwarzschild de-Sitter black hole shadow with respect to the position of a co-moving observer ($\frac{r_0}{M}$). We set the plasma parameter $\frac{k}{M}=0.2$. ]{\includegraphics[width=\textwidth]{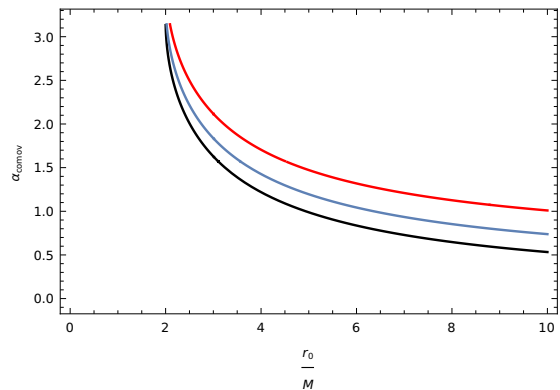}}\label{xx2}
\end{minipage}
\caption{Variation in the angular size of the shadow with respect to the position of the co-moving observer $\frac{r_0}{M}$. The plots are shown for different values of Hubble constant $H_0 M$ as 0.01 (black), 0.05 (violet) and 0.10 (red).}
\label{84}
\end{figure}
\noindent Using the expression of the comoving velocity of the observer, we can determine the angular shadow size with respect to a comoving observer. This we show in Fig.\ref{84}. The left plot represents the shadow without plasma scenario  and the right plot represents shadow in plasma background associated to the value $\frac{k}{M}=0.2$. Further, we also show the variation of the angular size of the black hole shadow for different values of the Hubbble constant $H_0$. From both the plots, for a fixed value of the observer position one can conclude that the shadow size increases with increase in the value of Hubble constant $H_0$ and the shadow size reduces to a finite value and never reaches zero size as the observer moves away from the black hole.

\begin{figure}[H]
\centering
\begin{minipage}[b]{0.4\textwidth}
\subfloat[\footnotesize Angular size ($\alpha_{comov}$) of the Schwarzschild de-Sitter black hole shadow with respect to the position of a co-moving observer ($\frac{r_0}{M}$) for various values of the plasma parameter $\frac{k}{M}$. ]{\includegraphics[width=\textwidth]{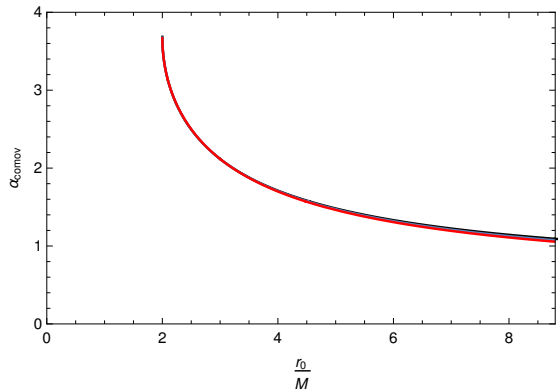}}\label{xx3}
\end{minipage}
\hspace{1.0cm}
\begin{minipage}[b]{0.4\textwidth}
\subfloat[\footnotesize  The zoomed version of the left image. ]{\includegraphics[width=\textwidth]{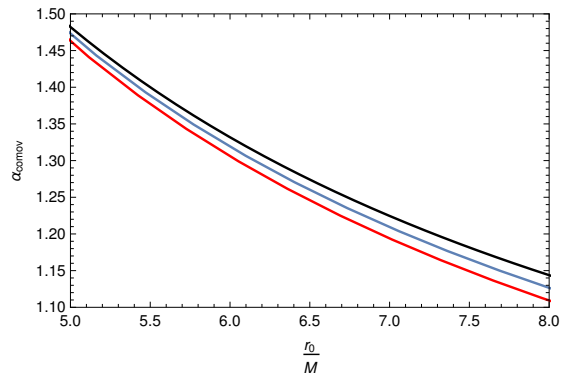}}\label{xx4}
\end{minipage}
\caption{Variation in the angular size of the shadow with respect to the position of the observer $\frac{r_0}{M}$. The plots are shown for different values of inhomogeneous plasma parameter $\frac{k}{M}$ as 0.0 (black), 0.2 (violet) and 0.4 (red)  with Hubble constant $H_0 M = 0.10$.}
\label{85}
\end{figure}
 \noindent In Fig.\ref{85}, we probe the effect of the plasma parameter on the angular shadow radius, as seen by a comoving observer. The plots have been shown for a fixed value of the Hubble constant $H_0 M =0.1$ and for inhomogeneous plasma parameter values $\frac{k}{M}= 0.0, 0.2$ and $0.4$. The plot in right is the zoomed version of the left one. We find that the increase of the value of the inhomogeneous plasma parameter $\frac{k}{M}$ reduces the angular size of the black hole shadow similar to that in case of a static observer.\\

\begin{figure}[H]
\centering
\begin{minipage}[b]{0.4\textwidth}
\subfloat[\footnotesize Silhouette of the black hole for different values of inhomogeneous plasma parameter $\frac{k}{M}$ with the observer positioned at $r_0 = 5M$. ]{\includegraphics[width=\textwidth]{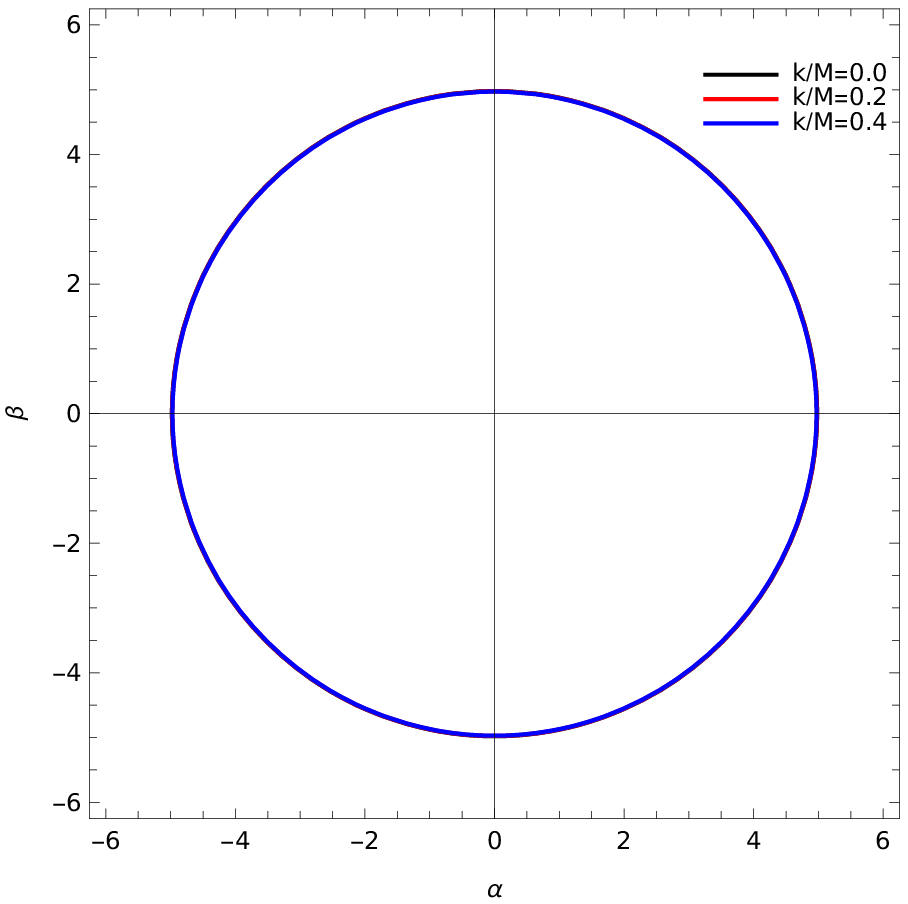}}\label{xx7}
\end{minipage}
\hspace{1.0cm}
\begin{minipage}[b]{0.4\textwidth}
\subfloat[\footnotesize  The zoomed version of the left image. ]{\includegraphics[width=\textwidth]{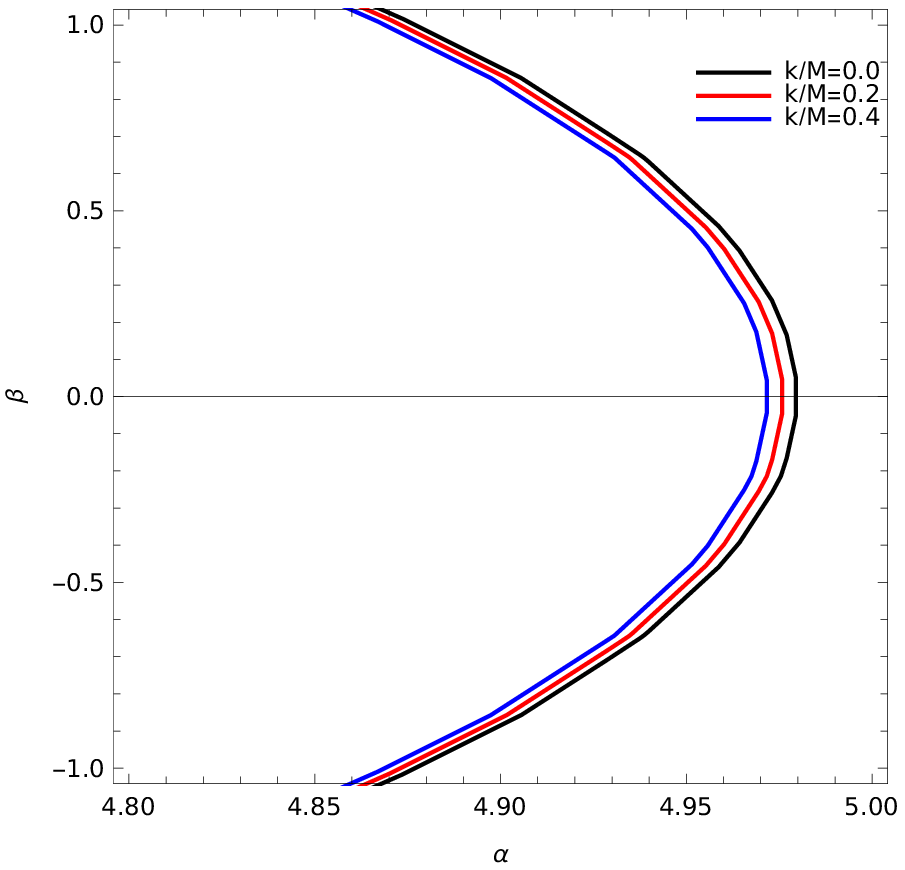}}\label{xx8}
\end{minipage}
\caption{Black hole shadow with variation of inhomogeneous plasma parameter $\frac{k}{M}$. The plots are shown for different values of inhomogeneous plasma parameter $\frac{k}{M}$ as 0.0 (black), 0.2 (red) and 0.4 (blue)  with Hubble constant $H_0 M = 0.10$. The observer is positioned at $r_0 = 5M.$}
\label{90}
\end{figure}

 \begin{figure}[H]
\centering
\begin{minipage}[b]{0.4\textwidth}
\subfloat[\footnotesize Silhouette of the black hole for different values of inhomogeneous plasma parameter $\frac{k}{M}$ with the observer positioned at $r_0 = 7M$.]{\includegraphics[width=\textwidth]{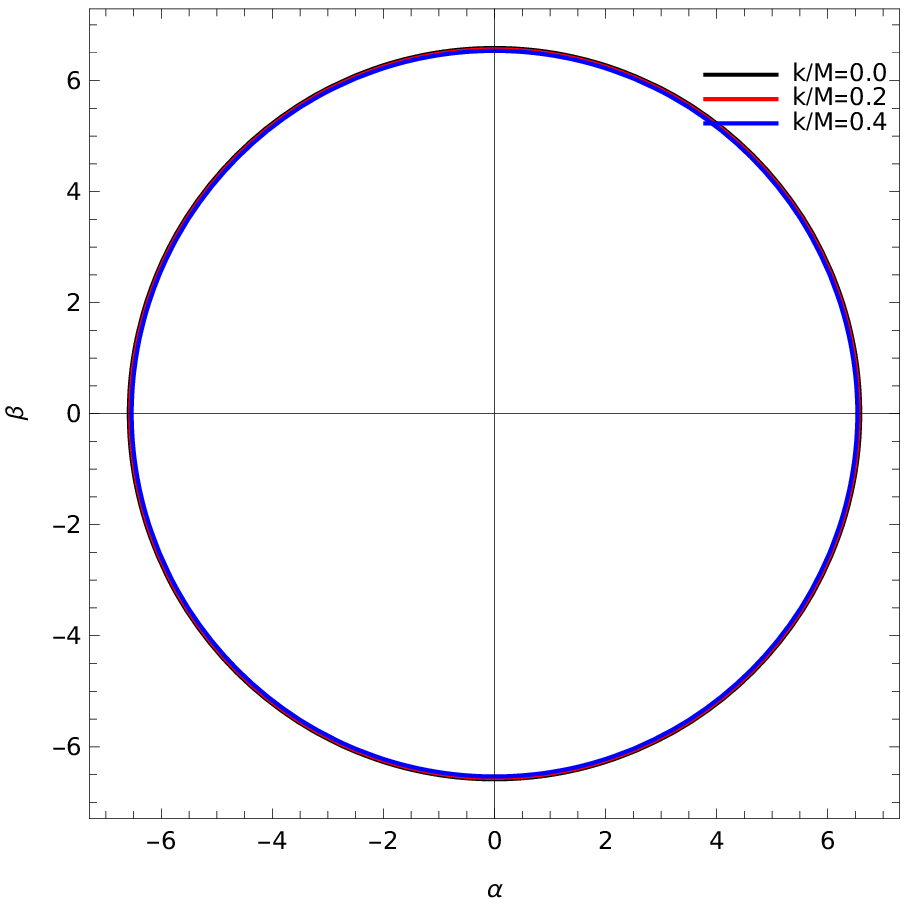}}\label{xx9}
\end{minipage}
\hspace{1.0cm}
\begin{minipage}[b]{0.4\textwidth}
\subfloat[\footnotesize  The zoomed version of the left image. ]{\includegraphics[width=\textwidth]{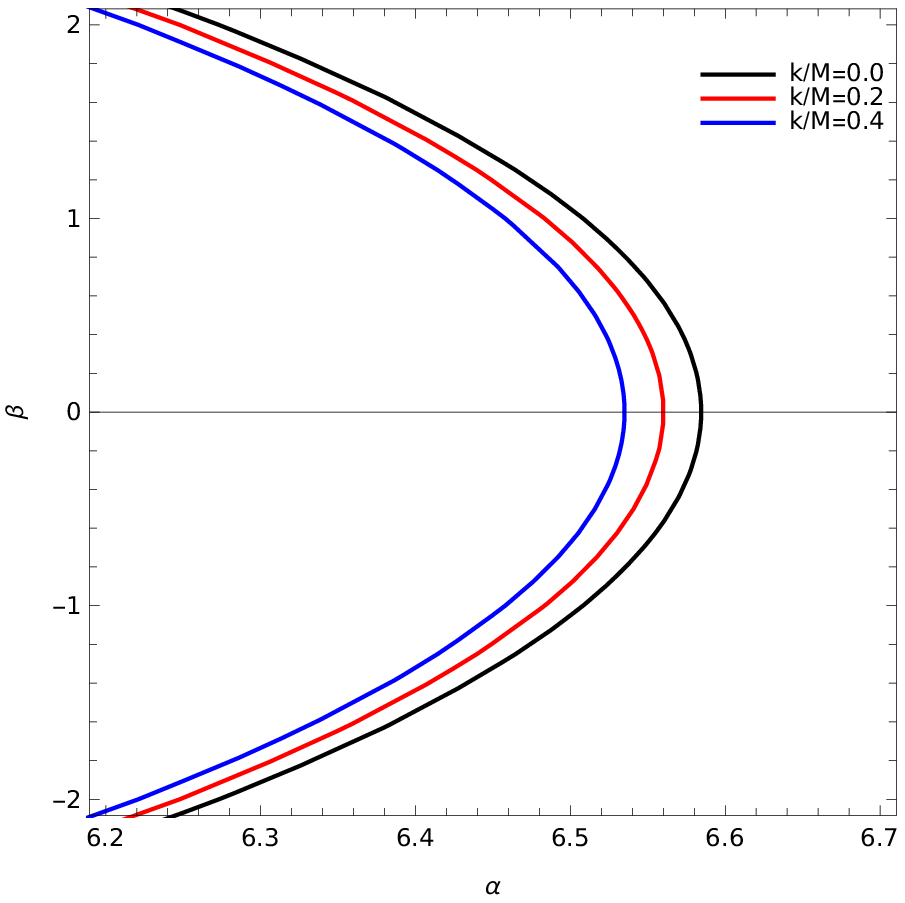}}\label{xx10}
\end{minipage}
\caption{Black hole shadow with variation of inhomogeneous plasma parameter $\frac{k}{M}$. The plots are shown for different values of inhomogeneous plasma parameter $\frac{k}{M}$ as 0.0 (black), 0.2 (red) and 0.4 (blue)  with Hubble constant $H_0 M = 0.10$. The observer is positioned at $r_0 = 7M.$}
\label{91}
\end{figure}
 
\noindent  In Fig.(s) \ref{90} and \ref{91}, we show the silhouette of the black hole shadow in the celestial plane. We have set the position of the observer at $r_0=5M$ in Fig. \ref{90} and $r_0 = 7M$ in Fig.\ref{91}. Since we are considering a spherically symmetric black hole, so the black hole shadow formed in the celestial plane ($\alpha$-$\beta$ plane) is circular. The radius of the shadow can be geometrically determined from Figure \ref{x1}, that is, $|R_s| =|r_0 \tan \alpha| \approx |r_0 \sin \alpha|$.  The shadow size gets reduced with increase in plasma parameter $\frac{k}{M}$ in both cases which can be clearly observed from the zoomed plots. Also, from the axes of the above Figures, it is clear that the shadow radius $R_s$ is larger in case of the observer positioned at $r_0=7M.$ Finally, Figures \ref{90} and \ref{91} also show the combined effect of plasma parameter $k$ and the cosmological constant $\Lambda$ on the shadow size, with $k$ having a de-magnification effect and $\Lambda$ having a magnification effect.

\section{Comparison with EHT observations}\label{sec5}

In this section, we compute the angular size of the black hole shadow and compare it with the observational results. We wish to compare the theoretical results with observations of supermassive black hole situated at the centre of M87 galaxy and our galaxy. We discuss them casewise.
\subsection{Case I: M87$^*$ supermassive black hole}

 The observed angular size of the black hole shadow at the centre of M87 galaxy is (42$\pm$3) $\mu$as  \cite{3}. Thus the angular shadow size varies in the range (39-45) $\mu$as. For the purpose of our calculation, we require the distance of the supermassive black hole from earth which is (16.8 $\pm$ 0.8) MPc  \cite{3}. The mass of the black hole as measured by the $EHT$ collaboration is ($6.5 \pm 0.7$) $\times 10^9$ \(M_\odot\)  \cite{3}. These are required to calculate the mass parameter $m$ given as $m=\frac{GM}{c^2}$ where $G$ is Newton's gravitational constant and $c$ is the velocity of light in vacuum having the standard values 6.674$\times$10$^{-11}$  m$^{3}$Kg$^{-1}$s$^{-2}$ and 3$\times$ 10$^{8}$ ms$^{-1}$ respectively.

 \noindent Also, our primary interest is in calculating the black hole shadow as observed by comoving observer. The comoving observer moves away with the cosmic expansion. In this case, we consider that the cosmic expansion is driven by the cosmological constant $\Lambda$ which is related to the Hubble constant as $\frac{\Lambda}{3}= \frac{H_0 ^2}{c^2}$. The angular shadow size as given in eq.(s) \eqref{20a} and \eqref{72} depend on the recession velocity of the comoving observer. The velocity of the comoving observer is given in \eqref{H} which is dependent on Hubble constant $H_0$. So, in order to determine the value of the angular black hole shadow size as measured by the comoving observer, we need the value of the Hubble's constant $H_0$. The value of $H_0$ as measured by the Planck Telescope is  ($67.4 \pm 0.5$) km s$^{-1}$MPc$^{-1}$ \cite{65}.

 \noindent Again, we calculate the shadow of a black hole surrounded by plasma medium. The plasma is assumed to be spherically symmetric and the plasma frequency $\omega_p$ and thereby the refractive index $n$ has only radial dependence. The refractive index takes the form $n=\sqrt{1-\frac{k}{r^h}f(r)}$. For homogeneous plasma, we set $h=0$ and for inhomogeneous plasma we consider the simplest case setting $h=1$. For these two cases we calculate the angular shadow size using \eqref{72}. From $\sin \alpha$, we calculate $\alpha$. This $\alpha$ gives the angular radius which is doubled to yield the angular diameter $2 \alpha$. This value is compared to the observational result to yield constraints on the value of the plasma parameter $k$.

\noindent Also, we convert the distance of black hole from earth $r_0$ as well as the photon sphere radius $r_p$ into metres. The value of $r_0$ is 5.184 $\times 10 ^{23}$ m. And the value of the photon radius for the Schwarzchild black hole both with and without cosmological constant $\Lambda$ is $3m$, that is, 2.892 $\times 10^{13}$ m. with mass parameter $m$ having value of 9.64 $\times 10^{12}$ m. Also, the cosmological horizon $r_C$ of the black hole spacetime is at 1.3736 $\times 10 ^{26}$ metre, where we have used the mass of the M87$^*$ supermassive black hole to be ($6.5 \pm 0.7$) $\times 10^9$ \(M_\odot\)  \cite{3} and the value of the Hubble constant as ($67.4 \pm 0.5$) km s$^{-1}$MPc$^{-1}$ \cite{65}. Thus, the observer is located well inside the horizon.

\begin{table}[h]
\centering
\begin{tabular}{|c|c|c|}
\multicolumn{2}{c}{}\\
\hline
$k$ & $\alpha_{stat}(\mu$as) \\
\hline
0.00 & 39.8643\\
\hline
0.05 & 40.7999\\
\hline
0.10 & 41.3061\\
\hline
0.15 & 42.1243\\
\hline
0.20 & 43.0210\\
\hline
0.25 & 44.0093\\
\hline
0.295 & 44.9907\\
\hline
0.296 & 45.0136\\
\hline
\end{tabular}
\hspace{3cm}
\begin{tabular}{|c|c|c|}
\multicolumn{2}{c}{}\\
\hline
$k$ & $\alpha_{comov}(\mu$as) \\
\hline
0.00 & 40.0150\\
\hline
0.05 & 40.9502\\
\hline
0.10 & 41.4542\\
\hline
0.15 & 42.2710\\
\hline
0.20 & 43.1633\\
\hline
0.25 & 44.1533\\
\hline
0.289 & 44.9971\\
\hline
0.29 & 45.0197\\
\hline
\end{tabular}
\caption{\footnotesize The Table shows the angular shadow size with change in plasma parameter $k$ for homogeneous plasma frequency $\omega_p ^2 = k = constant$. The left one is for static observer and right one is for comoving observer.}
\label{TabNew1}
\end{table}

\noindent Also we would like to point out that in the overall analysis, we have used a general expression of the velocity of the co-moving observer eq.\eqref{H}. For numerical analysis in cosmological scales, we need to consider the effective velocity given as \cite{x01}
\begin{equation}\label{09}
 v_{eff}=cz=H_0 D + v_p  \implies v_p = cz - H_0 D 
\end{equation}
where $v_{eff}$ gives the effective velocity which is given in terms of the redshift factor $z$. Also, $H_0$ gives the Hubble constant, $D$ is the distance of the observer from the black hole and $v_p$ is the peculiar velocity. Now  the redshift factor $z$ of the M87$^*$ black hole with respect to earth is 0.00428 \cite{x02}, the Hubble's constant $H_0$ is 67.4 km s$^{-1}$MPc$^{-1}$ \cite{65}  and $D$ is 16.8 $Mpc$  \cite{3}. Using these values, we obtain the peculiar velocity $v_p$ to be 178.64 km s$^{-1}$. However, the velocity due to cosmic expansion is 1105.36 km s$^{-1}$ which can be obtained using eq.\eqref{H}. In general we must have used $v_{eff}$ for our calculations, but the numerical results reveal that the co-moving velocity is quite large compared to the  peculiar velocity and hence we have neglected it in our work. Thus, our consideration of the earth as a co-moving frame neglecting the peculiar velocity gets valid.

\noindent Table \ref{TabNew1} shows the values of the angular size of the black hole shadow for different values of the plasma parameter $k$. The plasma medium considered here is of homogeneous nature with frequency $\omega_p ^2 = k  =$constant. The values of the angular shadow size varies within the range 39-45 $\mu$as. The values of the plasma parameter compatible with this range of the angular shadow size give us the bound on the parameter $k$. We find that the plasma parameter is bounded as $0 \leq k < 0.296$ for the shadow observed by the static observer and $0 \leq k < 0.29$ in case of the comoving observer. The left table in \ref{TabNew1} shows a list of values for $k$ compatible with the angular shadow size as observed by a static observer. On the other hand the right table shows the compatible range of values of $k$ with respect to a comoving observer. We find that the range of the values of plasma parameter $k$ are almost same in case of both the static and comoving observer.

\begin{table}[h]
\centering
\begin{tabular}{|c|c|c|}
\multicolumn{2}{c}{}\\
\hline
$\frac{k}{m}$ & $\alpha_{stat}(\mu$as) \\
\hline
0.00 & 39.8643\\
\hline
0.05 & 39.7534\\
\hline
0.10 & 39.6422\\
\hline
0.15 & 39.5307\\
\hline
0.20 & 39.4189\\
\hline
0.25 & 39.3067\\
\hline
0.30 & 39.1943\\
\hline
0.35 & 39.0815\\
\hline
0.386 & 39.0001\\
\hline
\end{tabular}
\hspace{3cm}
\begin{tabular}{|c|c|c|}
\multicolumn{2}{c}{}\\
\hline
$\frac{k}{m}$ & $\alpha_{comov}(\mu$as) \\
\hline
0.00 & 40.0150\\
\hline
0.05 & 39.7921\\
\hline
0.10 & 39.6801\\
\hline
0.15 & 39.6801\\
\hline
0.20 & 39.5679\\
\hline
0.25 & 39.4553\\
\hline
0.30 & 39.3424\\
\hline
0.35 & 39.2292\\
\hline
0.40 & 39.1157\\
\hline
0.4508 & 39.0000\\
\hline
\end{tabular}
\caption{\footnotesize The Table shows the angular shadow size with change in plasma parameter $\frac{k}{m}$ for inhomogeneous plasma frequency $\omega_p ^2= \omega ^2_p (r)=\frac{k}{r}$. The left one is for static observer and right one is for comoving observer.}
\label{TabNew2}
\end{table}
 
 \noindent The above Table \ref{TabNew2} shows the values of the inhomogeneous plasma parameter $\frac{k}{m}$ compatible with the values of the angular shadow size lying within the range 39-45 $\mu$as. The Table is constructed by considering the inhomogeneous plasma frequency of the form $\omega_p ^2 (r)=\frac{k}{r}$. The left table in \ref{TabNew2} is for the angular shadow size as observed by a static observer and that in the right is for comoving observer. The bound on the plasma parameter $\frac{k}{m}$ is given as $0 \leq \frac{k}{m}\leq 0.386$ for the static observer and $0 \leq \frac{k}{m} \leq 0.4508$ in case of comoving observer. The ranges of the value of plasma parameter $\frac{k}{m}$ is quite different for static and comoving observers. The range is extended more in case of comoving observer than that for the static observer.\\

\subsection{Case II: Sgr A$^*$ supermassive black hole}
We wish to constraint the plasma parameter in case of Sgr A$^*$ supermassive black hole. Such constraining of other types of parameters have been done in case of Sgr A$^*$ supermassive black hole in \cite{s1}-\cite{s4}. We want to carry out the analysis by comparing the theoretical and observed values of angular size of the black hole shadow. The observed value of the angular size of the black hole shadow is (48.7 $\pm$ 7) $\mu$as \cite{66}. In order to perform the numerical analysis, we require the black hole mass, which is  $ 4^{+1.1} _{-0.6}\times 10^6 $ \(M_\odot\) \cite{66}. The black hole is at a distance $(8277 \pm 9 \pm 33) pc$ \cite{67} from earth. We use the value of the black hole mass as $ 4\times 10^6 $ \(M_\odot\) and distance to the black hole as 8277 pc for our calculations. The angular size of the black hole shadow varies in the range (41.7 - 55.7) $\mu$as.\\ 

\begin{table}[h]
	\centering
	\begin{tabular}{|c|c|c|}
		\multicolumn{2}{c}{}\\
		\hline
		$k$ & $\alpha_{stat}(\mu$as) \\
		\hline
		0.000 & 49.6868\\
		\hline
		0.050 & 50.5446\\
		\hline
		0.100 & 51.4799\\
		\hline
		0.150 & 52.4996\\
		\hline
		0.200 & 53.6172\\
		\hline
		0.250 & 54.8490\\
		\hline
		0.281 & 55.6789\\
		\hline
		0.282 & 55.7066\\
		\hline
	\end{tabular}
	\hspace{3cm}
	\begin{tabular}{|c|c|c|}
		\multicolumn{2}{c}{}\\
		\hline
		$k$ & $\alpha_{comov}(\mu$as) \\
		\hline
		0.000 & 49.6869\\
		\hline
		0.050 & 50.5447\\
		\hline
		0.100 & 51.4800\\
		\hline
		0.150 & 52.4997\\
		\hline
		0.200 & 53.6173\\
		\hline
		0.250 & 54.8491\\
		\hline
		0.281 & 55.6790\\
		\hline
		0.282 & 55.7067\\
		\hline
	\end{tabular}
	\caption{\footnotesize The Table shows the angular shadow size with change in plasma parameter $k$ for homogeneous plasma frequency $\omega_p ^2 = k = constant$. The left one is for static observer and right one is for comoving observer.}
	\label{TabNew3}
\end{table}

\noindent We are interested in carrying out the analysis for both homogeneous and inhomogeneous plasma distribution. The refractive index in case of homogeneous plasma takes the form $n=\sqrt{1-kf(r)}$ with $k$ being the plasma parameter and $f(r)$ being the lapse function. Table \ref{TabNew3} shows the angular size of the black hole shadow measured by both static and co-moving observers. We present those values of the angular shadow which lie in the desired range $(41.7 - 55.7) \mu as$ along with the compatible values of the plasma parameter $k$. We find that the plasma parameter is bounded as $0 \leq k < 0.282$ for both static and co-moving observer. Also, we notice something significant. The angular size of the black hole shadow is same for both static and co-moving observer. This is due to the fact that the supermassive black hole Sgr A$^*$ and the observer in this case (earth) belongs to the same galactic frame.\\

\begin{table}[h]
	\centering
	\begin{tabular}{|c|c|c|}
		\multicolumn{2}{c}{}\\
		\hline
		$\frac{k}{m}$ & $\alpha_{stat}(\mu$as) \\
		\hline
		0.00 & 49.6868\\
		\hline
		0.50 & 48.2802\\
		\hline
		1.00 & 46.8166\\
		\hline
		1.50 & 45.2895\\
		\hline
		2.00 & 43.6907\\
		\hline
		2.50 & 42.0102\\
		\hline
		2.58 & 41.7330\\
		\hline
	\end{tabular}
	\hspace{3cm}
	\begin{tabular}{|c|c|c|}
		\multicolumn{2}{c}{}\\
		\hline
		$\frac{k}{m}$ & $\alpha_{comov}(\mu$as) \\
		\hline
		0.00 & 49.6869\\
		\hline
		0.50 & 48.2802\\
		\hline
		1.00 & 46.8167\\
		\hline
		1.50 & 45.2896\\
		\hline
		2.00 & 43.6907\\
		\hline
		2.50 & 42.0103\\
		\hline
		2.58 & 41.7331\\
		\hline
	\end{tabular}
	\caption{\footnotesize The Table shows the angular shadow size with change in plasma parameter $\frac{k}{m}$ for inhomogeneous plasma frequency $\omega_p ^2 = \frac{k}{r}$. The left one is for static observer and right one is for comoving observer.}
	\label{TabNew4}
\end{table}

\noindent Then we try to constraint the inhomogeneous plasma parameter $\frac{k}{m}$ using the angular size of black hole shadow. The refractive index in case of inhomogeneous plasma takes the form $n=\sqrt{1-\frac{k}{r}f(r)}$. Table \ref{TabNew4} shows the values of angular size of the black hole shadow and the compatible values of the plasma parameter $\frac{k}{m}$. We find that the plasma parameter $\frac{k}{m}$ is bounded as $0 \leq \frac{k}{m} \leq 2.58 $ in case of  both the static and the co-moving observer. Similar to the previous case, we find that the black hole shadow size is same for both static and co-moving observers.

\section{Conclusion}\label{sec6}
We now summarize our findings and conclude. In this work, we have performed the analysis of black hole shadow in presence of plasma. To be precise, we have calculated the angular shadow radius of a spherically symmetric black hole in presence of inhomogeneous plasma, as seen by both static and comoving observer. We start by considering a spherically symmetric metric black hole metric in (3+1) dimensions with arbitrary lapse function $f(r)$ and then consider lapse function of the Kottler or Schwarzchild de-Sitter black hole spacetime, having a positive cosmological constant $\Lambda>0$. This $\Lambda$ is responsible for the expansion of the universe. The important and unique results in our work is the effect of plasma in the black hole shadow with $\Lambda > 0$. We have evaluated the general equation for determining the photon sphere radius $r_p$ which is valid for both homogeneous and inhomogeneous plasma background. The equations are non-trivial and can be solved only numerically. Then we have shown the plots for angular shadow size for Schwarzschild and Schwarzschild de-Sitter black hole with variation in plasma. Here, we found that with increase in the inhomogeneous plasma parameter $k$, the shadow size reduces. We have also shown the effect of plasma on the photon sphere radius $r_p$. Later, we have shown the effect of cosmological constant $\Lambda$ on the angular shadow size. We observe that at the position of the observer which is well outside the photon sphere, the angular shadow size reduces with increase in $\Lambda$. The effect remains the same even in presence of plasma parameter.

\noindent  The graphical analysis reveal many interesting results. We observe that the angular shadow size varies in the range $\pi$ to $0$. The shadow size is $\pi$ at the black hole horizon and goes down to zero at infinity for general asymptotically flat black holes. In case of asymptotically de-Sitter black holes, the shadow goes to zero size at the cosmological horizon $r_C$. The angular shadow for co-moving observer is related to that of the static observer. The interesting feature is that in case of co-moving observer, there is no cosmological horizon and the observer can reach upto infinity. Further, in this case the shadow size does not go to zero and always remains finite. Since the observer is moving with the expansion, hence he never feels any horizon as such. \\
\noindent Then we plot and analysed the shadow size by considering plasma effects. We observed that plasma impacts the shadow size greatly. In general, practically an observer can exist outside the photon sphere $r_p$. From his perspective the shadow size changes with variation in plasma parameter $k$. In case of homogeneous plasma, we find that the shadow size increases with increase in plasma for both static and co-moving observer. On the other hand, we observe that the shadow size decreases with the increase in the value of the plasma parameter $k$ for inhomogeneous plasma. We have shown the plots only for inhomogeneous plasma. The new results in our work is the analysis of black hole shadow with respect to both static and comoving observer in presence of plasma.
 
\noindent Finally, we have compared the angular shadow size with the observational results of M87$^*$ and Sgr A$^*$ data. Comparing our results with the observed angular shadow size, we find bounds on the plasma parameter $k$. We find that for homogeneous plasma, $k$ has an upper bound of $k < 0.296$ for static observer and $k < 0.29$ for co-moving observer. In case of inhomogeneous plasma, we find that $\frac{k}{m}$ has an upper bound of $\frac{k}{m} \leq 0.386$ for static and $\frac{k}{m} \leq 0.4508$ for co-moving observer. These bounds are for M87$^*$ black hole. In case of Sgr A$^*$, we find that the angular shadow size is the same both for static and co-moving observers. This is due to the fact that the supermassive black hole Sgr A$^*$ and the observer in this case (earth) belongs to the same galactic frame. The bounds on the homogeneous and inhomogeneous plasma are given as $0 \leq k < 0.282$ and $0 \leq \frac{k}{m} \leq 2.58$ respectively. The constraints are true for both static and co-moving observers. We would like to mention that a more accurate estimation of the constraints on the plasma parameter $k$ would emerge from taking into consideration the spin parameter for rotating black holes. We wish to carry out such an analysis in future.

\appendix
\section*{Appendix: \\
Determination of the velocity of the comoving observer with respect to the static observer}\label{AppA}
\noindent In \textbf{Step I}, one considers the asymptotically flat version of the original metric eq.\eqref{m1} by dropping the cosmological constant, that is, by setting $\Lambda=0$. This gives
\begin{equation}\label{02}
ds_{\Lambda=0}^2= - f_0(r)dt^2 + \frac{1}{f_0(r)}dr^2 + r^2\Big(d\theta^2 + \sin^2 \theta d\phi^2\Big)
\end{equation}
where $f_0 (r) = f(r)\Big|_{\Lambda=0}$.

\noindent In \textbf{Step II},  one writes the metric in isotropic coordinates as
\begin{equation}\label{03}
ds_{\Lambda=0}^2= -  k(\tilde{r})dt^2 + g^2(\tilde{r})\Bigg(d\tilde{r}^2 + \tilde{r}^2\Big(d\theta^2 + \sin^2 \theta d\phi^2\Big)\Bigg)~.
\end{equation}
This eq.\eqref{03} can be compared with eq.\eqref{02} to get a relation between $r$ and $\tilde{r}$ given as
\begin{equation}
dr=\frac{r}{\tilde{r}}\sqrt{f_0(r)}d\tilde{r}~.
\end{equation}
The above relation is then used to replace $r$ in terms of $\tilde{r}$ in the original spherically symmetric metric given in eq.\eqref{m1}.\\

\noindent In \textbf{Step III},  one can make a transformation from $twiddled$ to the co-moving coordinate ($r_c, t_c$) as $\tilde{r}= r_c ~a (t_c)$ \cite{41} where, $r_c$ is the radial coordinate in the co-moving frame and $a$ is the scale factor of the universe which is the function of time ($t_c$) in the co-moving frame. This will lead us to a differential form of $d \tilde{r}$ in terms of $dr_c$ and $dt_c$. Similarly, one can assume that $dt$ is of the form
\begin{equation}\label{E2}
dt=B(t_c,r_c)dt_c + C(t_c,r_c)dr_c~.
\end{equation}
These coefficients will be determined by the next steps of the algorithm. Also due to the spherical symmetry (even though we go from original $\to$ twiddled $\to$ co-moving frame), $\theta$ and $\phi$ parts remain intact.\\

\noindent In \textbf{Step IV},  one can rewrite the metric in terms of $r_c, t_c, \theta, \phi$ coordinates as
\begin{eqnarray}
ds^2 &=& -\Bigg[ \frac{B^2}{f(r)} -\frac{f_0(r) r^2}{f (r)}\Big(\frac{\dot{a}}{a}\Big)^2 \Bigg]dt_c ^2 + \Bigg[ \frac{f_0(r) r^2}{f (r)r_c ^2} - \frac{C^2}{f(r)}\Bigg]dr_c ^2 + 2\Bigg[\frac{f_0(r) r^2}{f(r)r_c ^2}\frac{\dot{a}}{a} - \frac{BC}{f(r)}\Bigg]dt_c dr_c\nonumber\\
&&+ g^2(\tilde{r})\tilde{r}^2\Big(d\theta^2 + \sin^2 \theta d\phi^2\Big)~.
\end{eqnarray}
Imposing the condition that the final form of the metric is
isotropic and diagonal, we get
\begin{equation}
B=r\frac{\dot{a}}{a}\frac{f_0 (r)}{f(r)}\sqrt{f(r)}\Big[\frac{f_0(r)}{f(r)}-1\Big]^{-1/2}~~;~~ C=\frac{r}{r_c }\sqrt{f(r)}\Big[\frac{f_0(r)}{f(r)}-1\Big]^{1/2}~.
\end{equation}
Thus, one determines $dt$ and $dr$ in terms of $dt_c$ and $dr_c$. Using this relation, one readily determines $dt_c$ and $dr_c$ in terms of $dt$ and $dr$ which in turn gives $\partial_{t_c}$ and $\partial_{r_c}$. Finally using the normalisation condition $ g_{\mu \nu}U^{\mu}U^{\nu} =-1$ and also the relation between the four velocity of the co-moving observer with respect to the static observer as
\begin{equation}\label{210}
g_{\mu \nu}U^{\mu} _s U^{\nu}_c =-\frac{1}{\sqrt{1-v^2}}
\end{equation}
one can determine the velocity of the co-moving frame with respect to the static frame.

\noindent Using the above algorithm the velocity ($v$) of the comoving observer takes the form \cite{1}
\begin{equation}
|\Vec{v}|=\sqrt{1-\frac{f(r)}{f_0(r)}}.
\end{equation}
$f(r)$ is the lapse function with cosmological constant ($\Lambda$), whereas $f_0(r)$ is devoid of that term. Besides $\Lambda$ appears in the lapse function as $\frac{\Lambda}{3}r^2= \frac{H_0 ^2}{c^2}r^2$. For $c=1$, we get $\frac{\Lambda}{3}r^2= H_0 ^2 r^2$. Thus, the velocity of the co-moving observer becomes
\begin{equation}
|\Vec{v}|=\frac{H_0 r}{\sqrt{f_0(r)}}.
\end{equation}

\section*{Acknowledgments}

A.D. would like to acknowledge the support of S.N. Bose National Centre for Basic Sciences for Senior Research Fellowship. A.S. acknowledges the financial support by Council of Scientific and Industrial Research (CSIR, Govt. of India). The authors would like to thank the anonymous referee for useful suggestions and comments.

\end{document}